\documentclass
[%
reprint,
 amsmath,amssymb,aps,
]{revtex4-1}

\usepackage{graphicx}
\usepackage{dcolumn}
\usepackage{bm}
\usepackage{hyperref}
\usepackage{csquotes}
\usepackage{xcolor}
\usepackage{pgfplots}
\usepackage[normalem]{ulem}

\begin{document}

\title{Majoron Dark Energy via Freezing Induced by Quantum Coherence}

\author{Keunsu Cheon}
\affiliation{Graduate School Department of Physics, Daejin University, Pocheon 11159, Korea}

\author{Sin Kyu Kang}
\affiliation{Seoul National University of Science and Technology, Seoul 01811, Korea}

\author{Jungjai Lee}
\email{jjlee@daejin.ac.kr}
\affiliation{Graduate School Department of Physics, Daejin University, Pocheon 11159, Korea}
\affiliation{School of Physics, Korea Institute for Advanced Study, Seoul 02455, Korea}

\date{\today}

\begin{abstract}
We propose a nonequilibrium mechanism for Majoron dark energy in which the late-time freezing of a physical Majoron is induced by quantum coherence in a hidden pseudo-Dirac sterile fermion reservoir. The evolving Majoron background derivatively couples to the hidden pseudo-Dirac number current and drives a lagged reservoir response with a finite memory time. 
In the short-memory regime, the causal response kernel reduces to \(\dot X+\Gamma_{\rm PD}X=\beta\ddot\phi\). The leading linear-response matching \(Q=\alpha X\) then yields an effective scalar equation containing the exchange structure \(q_{\rm exch}\ddot\phi/\dot\phi\). We show that this term can dynamically suppress the Majoron velocity and sustain a response-dominated freezing branch even when the intrinsic Majoron mass is larger than the present Hubble scale. The microscopic origin of the lag variable is identified with the phase-lagged off-diagonal coherence of the hidden pseudo-Dirac ensemble, while the response strength is controlled by a response-weighted hidden density rather than by an independent gravitating component. The resulting state is a metastable nonequilibrium frozen phase with \(w_\phi\simeq -1\), rather than an exactly static cosmological constant.
\end{abstract}

\maketitle

\section{Introduction}

The observed accelerated expansion of the Universe strongly suggests the existence of a dominant dark-energy component in the present cosmological epoch~\cite{Riess1998,Perlmutter1999,Planck2018}. In the standard $\Lambda$CDM framework, the late-time acceleration is attributed to a cosmological constant. This description is phenomenologically successful at the background level and remains the simplest effective model of the observed cosmic acceleration. Nevertheless, the microscopic origin of the dark-energy scale remains one of the deepest unresolved problems in modern theoretical physics. If the observed dark-energy density is interpreted as a true vacuum energy, its magnitude is extraordinarily small compared with natural particle-physics scales. If it is instead interpreted as an effective late-time component, one must explain why this component remains nearly vacuum-like over cosmological timescales.

Recent observations have also motivated renewed interest in dynamical dark-energy scenarios. In particular, baryon acoustic oscillation measurements and their combinations with cosmic microwave background and supernova data have sharpened constraints on the late-time expansion history~\cite{DESI2024a,DESI2024b,DESI2025DR2BAO,DESI2025DE}. These developments do not by themselves establish dynamical dark energy, and their interpretation depends on data combinations, parameterizations, and consistency assumptions. They do, however, strengthen the motivation for studying theoretically controlled mechanisms in which vacuum-like behavior may arise dynamically rather than being imposed as an exactly constant vacuum energy.

A widely studied alternative to a cosmological constant is scalar-field dark 
energy. In quintessence-type models, a scalar field evolves along a sufficiently 
flat potential and can produce an equation of state close to $w\simeq -1$ at late 
times~\cite{RatraPeebles1988,Wetterich1988,Caldwell1998}. Depending on the 
potential and initial conditions, scalar dark energy can exhibit tracking, 
thawing, or freezing behavior~\cite{Zlatev1999,CaldwellLinder2005}. In 
conventional freezing scenarios, the scalar velocity decreases as the field 
evolves toward a region of the potential where its kinetic energy becomes small 
compared with its potential energy. However, such models usually require the 
relevant scalar mass scale to be of order the present Hubble scale, or smaller, 
$m_\phi\lesssim H_0$, in order to avoid the onset of coherent oscillations in the 
late Universe. Once Hubble friction becomes weaker than the intrinsic scalar 
curvature scale, $H\lesssim m_\phi$, a coherently displaced scalar field generally 
begins to oscillate around the minimum of its potential. For a quadratic minimum, $V(\phi)\simeq m_\phi^2\phi^2/2$, the cycle-averaged kinetic energy $K\equiv\dot\phi^2/2$ and potential energy satisfy $\langle K\rangle\simeq\langle V\rangle$. Hence the positive 
pressure from the kinetic term cancels the negative pressure from the potential term on average, giving $\langle p_\phi\rangle\simeq0$ and $\langle w_\phi\rangle\simeq0$. The scalar condensate then redshifts as pressureless matter, $\langle\rho_\phi\rangle\propto a^{-3}$, instead of remaining vacuum-like.

This requirement is particularly restrictive for particle-physics motivated scalar fields. Pseudo-Nambu-Goldstone bosons (pNGBs) are attractive dark-energy candidates because approximate shift symmetries can protect small masses and flat potentials~\cite{Frieman1995}. Majoron models associated with spontaneously broken lepton number were introduced in
Refs.~\cite{Chikashige1981,Gelmini1981}. A gauged \(U(1)_{B-L}\) realization with an additional singlet scalar, motivated by Planck-scale breaking of global symmetries, was studied in Ref.~\cite{Babu1993}. Recent work has also revisited the cosmological role of Majoron-like
pNGBs, including electroweak-instanton effects, domain-wall issues, and
connections to axion-like phenomenology
\cite{Berbig2026MajoronWalls,Liang2025MajoronAxion}.
However, if a Majoron is to play the role of dark energy in the conventional slow-roll way, its effective curvature scale must still be extremely small at late times. This motivates the question addressed in this paper: can a Majoron remain dark-energy-like not because its potential is exceptionally flat, but because its motion is dynamically suppressed by the delayed response of a hidden nonequilibrium medium?

We propose that such a mechanism can be realized if the physical Majoron is coupled derivatively to a hidden pseudo-Dirac sterile fermion reservoir. The hidden reservoir is not assumed to be an ordinary thermal bath. Rather, it is a coherent hidden medium with finite memory and relaxation timescales. As the homogeneous Majoron background evolves, it drives the hidden pseudo-Dirac current. Because the reservoir cannot instantaneously reorganize itself, its response develops a finite lag. After coarse graining over the hidden degrees of freedom, this finite-memory response is described by a causal retarded kernel and, in the short-memory limit, by a local lag equation~\cite{Kubo1957,Breuer2002}.

The central effective structure is a causal response relation $X(t)=\int_{-\infty}^{t}dt'K(t-t')F(t')$, where $X$ is the collective lag variable and $F$ is the Majoron-induced drive. For an exponential short-memory kernel, this relation reduces to the local lag equation $\dot X+\Gamma_{\rm PD}X=\beta F$. Since the derivative Majoron interaction makes the instantaneous hidden-sector perturbation proportional to $\dot\phi$, the lag is sourced by the time variation of this perturbation, so that $F(t)=\ddot\phi(t)$. The local lag equation then becomes $\dot X+\Gamma_{\rm PD}X=\beta\ddot\phi$. The physical energy transfer between the Majoron sector and the hidden reservoir is matched as $Q=\alpha X$, and in the Markovian regime this gives $Q\simeq q_{\rm exch}\ddot\phi$. Inserting this energy transfer into the scalar energy-balance equation yields an effective scalar equation containing a response-induced exchange structure proportional to $q_{\rm exch}\ddot\phi/\dot\phi$. This term is not introduced as an ad hoc local friction force. It is the Markovian remnant of the retarded response of the hidden pseudo-Dirac reservoir.

The resulting freezing is characterized by the response-dominated hierarchy $|\dot\phi|\ll q_{\rm exch}$ and $3H|\dot\phi|\ll |V_\phi|$. The form of these freezing hierarchies is motivated by the general freezing criteria discussed in Ref.~\cite{Simpson2018}. 
In the present work, however, \(q_{\rm exch}\) denotes the exchange coefficient generated by the hidden pseudo-Dirac reservoir response. In this regime, the Majoron velocity is dynamically suppressed by the hidden reservoir response rather than by ordinary Hubble friction alone. The scalar kinetic energy remains small, and the equation of state approaches $w_\phi\simeq -1$. The resulting state is a metastable nonequilibrium frozen phase, not an exactly static de Sitter vacuum.

This paper is organized as follows. In Sec.~II we define the effective Majoron sector and the hidden pseudo-Dirac reservoir. In Sec.~III we derive the microscopic origin of the collective lag variable from the off-diagonal coherence of the hidden pseudo-Dirac ensemble. In Sec.~IV we construct the retarded-response equation and derive the effective freezing equation. In Sec.~V we discuss the cosmological freezing branch, the exchange coefficient, the Markovian consistency window, and the finite-memory non-Markovian regime. Section~VI summarizes the results and outlines future directions. Technical details are collected in the Appendices.

\section{Majoron and hidden pseudo-Dirac reservoir}

\subsection{Physical Majoron sector}

We consider a physical Majoron associated with the spontaneous breaking of a gauged $U(1)_{B-L}$ symmetry. Since the symmetry is gauged, one phase direction is eaten by the massive gauge boson, while an orthogonal compact direction remains as a physical low-energy scalar. 

For two complex singlet fields \(\Phi_i\), with \(i=1,2\), carrying \(U(1)_{B-L}\) charges \(q_i\), we write
\begin{equation}
\Phi_i(x)
=
\frac{1}{\sqrt{2}}
\left(v_i+\rho_i(x)\right)e^{i\theta_i(x)},
\qquad
\theta_i\sim\theta_i+2\pi .
\label{eq:Phi-param}
\end{equation}
The renormalizable scalar potential fixes the symmetry-breaking vacuum, \(\langle \Phi_i\rangle=v_i/\sqrt2\), and determines the radial-mode mass spectrum. At the renormalizable level, however, it is independent of the uneaten phase direction. Thus the physical Majoron remains massless before explicit breaking. 

A small periodic potential for this physical Majoron is generated by a gauge-invariant Planck-suppressed operator, whose projection onto the uneaten phase direction is discussed in Appendix~\ref{app:scalar_potential_majoron}.
At energies well below the radial-mode masses, the heavy modes $\rho_i$ are integrated out, and the low-energy dynamics is described by the compact phase degrees of freedom $\theta_i$.
The corresponding phase kinetic term is
\begin{equation}
{\cal L}_{\rm phase}
=
\frac12
\sum_{i=1}^{2}
v_i^2
\left(\partial_\mu\theta_i-q_iB_\mu\right)
\left(\partial^\mu\theta_i-q_iB^\mu\right).
\label{eq:phase-lagrangian}
\end{equation}
It is understood that \(g_{B-L}\) has been absorbed into \(B_\mu\) in the phase kinetic term. The quantity \(M_0\) defined below therefore denotes the charge-weighted symmetry-breaking scale rather than the gauge-boson mass; with a canonically normalized gauge field, \(m_{B-L}=g_{B-L}M_0\).

The gauge-eaten direction is aligned with the charge vector, while the physical uneaten direction is orthogonal with respect to the kinetic metric. Defining
\begin{equation}
M_0=\sqrt{q_1^2v_1^2+q_2^2v_2^2},
\end{equation}
the canonically normalized uneaten phase direction may be written as
\begin{equation}
\phi=\frac{v_1v_2}{M_0}(q_2\theta_1-q_1\theta_2),
\end{equation}
which we refer to as the physical Majoron throughout this work.
The corresponding compact period is characterized by
\begin{equation}
f_{\rm eff}
=
\frac{v_1v_2\,{\rm gcd}(q_1,q_2)}
{\sqrt{q_1^2v_1^2+q_2^2v_2^2}} .
\end{equation}
This scale is the fundamental compact-period scale of the physical Majoron, not merely an operator-dependent coupling-suppression scale \cite{FraserReece2020}.

The benchmark charge assignment is chosen as
\begin{equation}
q_1=2,\qquad q_2=13.
\label{eq:q1q2}
\end{equation}
Its main role is to make the leading Planck-suppressed breaking operator
sufficiently high-dimensional so that, for the values of $v_1$ and
$v_2$ used below, the induced Majoron potential naturally falls near
the observed dark-energy scale. More explicitly, the charges are coprime, so that the uneaten Majoron
direction has a fundamental compact period without an additional common
charge factor. Indeed, a monomial \(\Phi_1^m(\Phi_2^\dagger)^n\) is gauge invariant only if
\[
mq_1-nq_2=0 .
\]
For \(q_1=2\) and \(q_2=13\), the smallest positive integer solution is
\(m=13\) and \(n=2\). Thus the leading phase-sensitive operator is
\begin{equation}
{V}_{\rm br}
=
-\kappa\frac{\Phi_1^{13}(\Phi_2^\dagger)^2}{M_{\rm Pl}^{11}}
+{\rm h.c.},
\label{eq:breaking_operator}
\end{equation}
where \(M_{\rm Pl}\) denotes the unreduced Planck scale. The renormalizable
scalar potential therefore leaves the physical Majoron direction massless,
while the Planck-suppressed operator in Eq.~\eqref{eq:breaking_operator}
generates a small periodic potential.

For the charge assignment $\rm{gcd}(2,13)=1$ in Eq.~\eqref{eq:q1q2}, the fundamental compact-period
scale becomes
\begin{equation}
f_{\rm eff}
=
\frac{v_1v_2}{\sqrt{4v_1^2+169v_2^2}} .
\label{eq:feff_benchmark}
\end{equation}
Projecting Eq.~\eqref{eq:breaking_operator} onto the physical Majoron direction
gives
\begin{equation}
V(\phi)
=
\Lambda_{\phi}^4
\left[
1-\cos\left(\frac{\phi}{f_{\rm eff}}+\delta\right)
\right],
\label{eq:majoron_potential}
\end{equation}
and near a minimum
\begin{equation}
m_{\phi}^2\simeq \frac{\Lambda_{\phi}^4}{f_{\rm eff}^2}.
\label{eq:mJ}
\end{equation}

The benchmark values of $v_1$ and $v_2$, together with the charge
assignment specified above, are chosen to illustrate that the same
$(B-L)$-breaking scalar sector can accommodate both an ordinary Type-I
seesaw scale and a dark-energy-scale Majoron potential. We take
\begin{align}
v_1 \simeq 2.2\times 10^{11}\,{\rm GeV},
\qquad
v_2 \simeq 1.0\times 10^{9}\,{\rm GeV}.
\label{eq:BM-scales}
\end{align}
The larger scale $v_1$ sets the mass scale of an ordinary Type-I seesaw
sector. For right-handed neutrinos with $B-L=-1$, the coupling
$y_M\Phi_1 N_R N_R$ gives $ M_R=y_M v_1/\sqrt{2}$. For \(y_M\sim{\cal O}(1)\), this gives
\(M_R\sim 10^{11}\,{\rm GeV}\), and light-neutrino masses of order
\(m_\nu\sim 0.05\,{\rm eV}\) can be obtained with a moderate Dirac Yukawa
coupling, \(y_D\sim 10^{-2}\). 

The lower scale \(v_2\) is chosen,
 together with the charge assignment of $q_1$ and $q_2$ in Eq.~(\ref{eq:q1q2}), so that the dimension-15 Planck-suppressed operator  in Eq.~(\ref{eq:breaking_operator}) yields a dark-energy-scale Majoron potential.
Inserting the values given in Eq.~(\ref{eq:BM-scales}) into Eq.~(\ref{eq:feff_benchmark}), one finds
\[
f_{\rm eff}\simeq 5.0\times 10^{8}\,{\rm GeV}.
\]
For dark-energy-scale potential energy density
\[
\rho_\phi\simeq \Lambda_{\phi}^4\simeq 2.6\times 10^{-11}\,{\rm eV}^4,
\]
one obtains
\[
m_{\phi}\sim 10^{-23}\,{\rm eV}.
\]

This value is much larger than the present Hubble scale, \(H_0\simeq 1.4\times 10^{-33}\,{\rm eV}\).
It is numerically close to the mass range often considered in ultralight
or fuzzy scalar dark matter models~\cite{Marsh2016,Hui2017}. In those
scenarios, however, the scalar is assumed to enter an oscillatory regime
and behaves as pressureless matter. By contrast, in the present framework
the hidden pseudo-Dirac reservoir suppresses the onset of coherent
Majoron oscillations, keeping the scalar potential-dominated with
\(w_\phi\simeq -1\). The benchmark mass should therefore be interpreted
not as a fuzzy-dark-matter choice, but as an illustration that
retarded-response freezing can maintain dark-energy-like behavior even
when \(m_{\phi}\gg H_0\).

Thus the intrinsic Majoron mass can be much larger than the present Hubble
scale, \(m_{\phi}\gg H_0\), while the potential energy remains of the observed
dark-energy order. The mechanism developed below is designed precisely for
this regime: the Majoron need not be frozen by requiring \(m_{\phi}\sim H_0\);
instead, its motion can be dynamically suppressed by the retarded response of
the hidden pseudo-Dirac reservoir.

\subsection{Hidden pseudo-Dirac reservoir}

The hidden pseudo-Dirac reservoir consists of two sterile Weyl fermions $N_h$ and $S_h$ forming a nearly degenerate pseudo-Dirac pair \cite{PseudoDirac1,PseudoDirac2}.
As derived in Appendix~\ref{app:hidden_pseudo_dirac_origin}, the hidden
pseudo-Dirac mass terms can be written as
\begin{equation}
{\cal L}_{\rm mass}^{h}
=
-m_N N_hS_h
-\frac12\mu_h N_hN_h
-\frac12\mu_h S_hS_h
+{\rm h.c.}
\label{eq:hidden-mass}
\end{equation}
Here \(m_N\) is the dominant Dirac mass scale, while \(\mu_h\ll m_N\) parametrizes the small Majorana splitting. In the minimal benchmark considered in this work, the hidden fields
\(N_h\) and \(S_h\) are distinct from the ordinary right-handed neutrinos
that participate in the Type-I seesaw mechanism. We take them to be
singlets under the gauged \(U(1)_{B-L}\),
\[
Q_{B-L}(N_h)=Q_{B-L}(S_h)=0 .
\]
This choice keeps the pseudo-Dirac reservoir as a genuinely hidden
response sector, separated from both the ordinary seesaw sector and direct
\(B-L\) gauge interactions. The derivative coupling between the physical
Majoron and the hidden pseudo-Dirac current is therefore not a minimal gauge
interaction implied by the \(B-L\) charge assignment. Rather, it should be
understood as a low-energy effective portal interaction between the pNGB-like
Majoron direction and a hidden sterile current. Such derivative couplings of
pNGB or axion-like fields to matter currents are standard in low-energy
effective descriptions of spontaneously broken symmetries \cite{Bauer2021ALPEFT}.
Related ALP-portal constructions
have also been used to connect axion-like degrees of freedom to dark
sectors \cite{Allen2024AxionPortal,DEramo2025AxionPortal}.

The symmetric choice of the two small Majorana entries in Eq.~\eqref{eq:hidden-mass}
is a benchmark simplification. It may be viewed as the limit selected by an
approximate hidden exchange symmetry \(N_h\leftrightarrow S_h\). This
assumption is not essential for the existence of a pseudo-Dirac reservoir;
a small asymmetric deformation would only split the two-state parameters and
can be treated perturbatively. The exchange-symmetric limit is adopted because
it reduces the hidden two-state system to a single Majorana splitting
parameter \(\mu_h\), making the coherence dynamics analytically transparent.
The limit \(\mu_h\to0\) restores the hidden pseudo-Dirac number symmetry, so
the small splitting is technically natural. For a nearly degenerate pseudo-Dirac pair, we denote by
\(E_p\equiv\sqrt{p^2+m_N^2}\) the common zeroth-order energy obtained
by neglecting the small Majorana splitting~\cite{Kayser1981,Beuthe2003}. Equivalently,
\(E_p\simeq \bar E_p\equiv(E_{1p}+E_{2p})/2\), up to corrections of
order \(\mu_h^2\). Therefore, to leading order in the pseudo-Dirac
splitting,
\begin{equation}
\Delta E_p
\equiv E_{1p}-E_{2p}
=
\frac{\Delta m_h^2}{2\bar E_p}
\simeq
\frac{\Delta m_h^2}{2E_p}
\simeq
\frac{2m_N\mu_h}{E_p}.
\end{equation}
This splitting controls the coherent relative phase evolution of the
two-state system. It is not generated by the Majoron motion; rather, it is an intrinsic microscopic scale of the hidden pseudo-Dirac pair.

The physical Majoron couples derivatively to the hidden pseudo-Dirac current,
\begin{equation}
{\cal L}_{\phi h}
=
\frac{\partial_\mu\phi}{f_{\rm eff}}J^\mu_{\rm PD}.
\end{equation}
For a homogeneous Majoron background, this contains
\begin{equation}
{\cal L}_{\phi h}
\supset
\frac{\dot\phi}{f_{\rm eff}}J^0_{\rm PD}.
\end{equation}
Thus the scalar velocity acts as a time-dependent drive on the hidden pseudo-Dirac ensemble. If the reservoir responded instantaneously, it would simply track the instantaneous value of the Majoron-induced perturbation. The essential point of the present mechanism is that the hidden reservoir has finite coherence and relaxation timescales, and hence responds with a delay.

The hidden pseudo-Dirac current is defined as the normal-ordered vector current
\begin{equation}
J^\mu_{\rm PD}
\equiv
:\bar N_D\gamma^\mu N_D: .
\label{eq:JPD-vector-current}
\end{equation}
Here \(N_D=(N_h,S_h^\dagger)^T\), with \(N_h\) and \(S_h\) denoting two-component Weyl field operators. The time component of the vector bilinear before normal ordering contains
\[
\bar N_D\gamma^0N_D
=
N_h^\dagger N_h+S_hS_h^\dagger .
\]
Using the equal-time fermionic anticommutation relation, the second term differs from \(-S_h^\dagger S_h\) by a vacuum contact term. Normal ordering removes this contact term, giving
\[
:\!S_hS_h^\dagger\!:
=
-S_h^\dagger S_h .
\]
Therefore the normal-ordered pseudo-Dirac number density is
\begin{equation}
J^0_{\rm PD}
=
:\bar N_D\gamma^0N_D:
=
N_h^\dagger N_h-S_h^\dagger S_h .
\label{eq:JPD-zero-weyl}
\end{equation}
In the pseudo-Dirac mass basis this pseudo-Dirac number density is an off-diagonal operator and is therefore controlled by the coherence between the two nearly degenerate states. This observation will be central in Sec.~III.

The hidden pseudo-Dirac fields are taken to be separate from the
right-handed neutrinos responsible for the observed active-neutrino
masses. The usual Type-I seesaw sector may be present, but it does not
supply the dominant response reservoir. The hidden fields form a
separate sterile reservoir whose active-hidden mixing is assumed to be
forbidden or sufficiently suppressed by gauge invariance and by an
approximate sterile-sector organizing symmetry. This separation allows
the response mechanism to be dominated by the hidden pseudo-Dirac
current rather than by the ordinary active-neutrino background.

\subsection{Cold nonrelativistic response regime}
Having isolated the hidden pseudo-Dirac reservoir from the ordinary neutrino
sector, we now specify the kinematic regime used for the leading response
estimate. We focus on a cold nonrelativistic hidden reservoir whose
response-weighted momentum support is concentrated around a representative
physical momentum \(p_*\) satisfying
\[
\frac{p_*^2}{m_N^2}\ll 1 .
\]
Hence, over the dominant support,
\begin{equation}
E_p\simeq m_N+\frac{p^2}{2m_N},
\qquad
\Delta E_p\simeq \frac{2m_N\mu_h}{E_p}\simeq 2\mu_h .
\label{eq:cold-nr-splitting}
\end{equation}

This cold-reservoir assumption reduces phase dispersion among
different momentum modes and allows smooth mode-dependent response
factors to be evaluated at \(p_\ast\), up to corrections controlled by the response-weighted momentum spread.

\section{Microscopic origin of the lag variable}

\subsection{Density matrix for the pseudo-Dirac pair}

We now describe how the collective lag variable arises from the microscopic coherence of the hidden pseudo-Dirac ensemble. For a fixed momentum mode $p$, the reduced density matrix in the pseudo-Dirac mass basis is
\begin{equation}
\rho_p
=
\begin{pmatrix}
\rho_{11}(p) & \rho_{12}(p)\\
\rho_{21}(p) & \rho_{22}(p)
\end{pmatrix},
\qquad
\rho_{21}(p)=\rho_{12}^*(p).
\end{equation}
The use of a momentum-dependent density matrix is standard in quantum
kinetic treatments of mixed fermionic ensembles
\cite{SiglRaffelt1993,Vlasenko2014,Blaschke2016}.
The diagonal entries determine occupation numbers and hence the physical hidden density $\rho_N$. The off-diagonal entry determines the coherence between the two pseudo-Dirac states and is the microscopic origin of the lagged current response discussed below.

In the two-state subspace spanned by the pseudo-Dirac mass eigenstates,
the common energy proportional to the identity does not affect the
commutator and may be dropped. With
\(\Delta E_p\equiv E_{1p}-E_{2p}\), we write
\begin{equation}
H_{0,p}
=
\frac{\Delta E_p}{2}\sigma_z .
\end{equation}
Here \(\sigma_z\) and \(\sigma_y\) denote Pauli matrices acting on the
two-dimensional pseudo-Dirac mass-eigenstate space
\((|1,p\rangle,|2,p\rangle)\). With this convention,
\(\sigma_z={\rm diag}(1,-1)\), so that
\(\dot\rho_{12}=-i\Delta E_p\rho_{12}\) in the absence of drive and
dissipation.

For each physical momentum mode \(p\), let
\({\cal H}_p\) denote the two-dimensional one-particle subspace spanned
by the two pseudo-Dirac mass eigenstates \(|1,p\rangle\) and
\(|2,p\rangle\). We define
\begin{equation}
P_p
=
|1,p\rangle\langle 1,p|
+
|2,p\rangle\langle 2,p|
\end{equation}
as the projector onto this fixed-\(p\) two-state subspace. In the phase
convention used here, the pseudo-Dirac number-density operator
projected onto this subspace is off diagonal,
\begin{equation}
P_pJ^0_{\rm PD}P_p
=
a_p\sigma_y ,
\end{equation}
where \(a_p\) is the corresponding current matrix element.

Since the homogeneous Majoron background gives
\({\cal L}_{\phi h}\supset \dot\phi\,J^0_{\rm PD}/f_{\rm eff}\), the
corresponding interaction Hamiltonian projected onto the same two-state
subspace is
\begin{equation}
H_{\phi,p}(t)
=
-\frac{\dot\phi(t)}{f_{\rm eff}}P_pJ^0_{\rm PD}P_p
=
-a_p\frac{\dot\phi(t)}{f_{\rm eff}}\sigma_y
\equiv
g_p(t)\sigma_y ,
\end{equation}
with
\begin{equation}
g_p(t)
\equiv
-a_p\frac{\dot\phi(t)}{f_{\rm eff}} .
\end{equation}
Thus the Majoron motion enters the two-state system as a Hamiltonian
drive, while relaxation and decoherence are encoded separately in the
dissipative term.

The evolution of the density matrix can be modeled as an open-system equation,
\begin{equation}
\dot\rho_p
=
-i[H_{0,p}+H_{\phi,p}(t),\rho_p]
+
{\cal D}[\rho_p].
\end{equation}
Here $H_{0,p}$ contains the pseudo-Dirac splitting, $H_{\phi,p}(t)$ is the Majoron-induced Hamiltonian drive, and ${\cal D}[\rho_p]$ is a Lindblad-type dissipator that parameterizes relaxation or decoherence of the hidden reservoir\footnote{For the minimal phenomenological treatment, the dissipator may be taken as a
pure-dephasing Lindblad term generated by
\[
L_{\rm deph}=\sqrt{\frac{\Gamma_{\rm PD}}{2}}\,\sigma_z .
\]
Then
\[
{\cal D}[\rho_p]
=
L_{\rm deph}\rho_p L_{\rm deph}^\dagger
-\frac12\left\{L_{\rm deph}^\dagger L_{\rm deph},\rho_p\right\},
\]
which gives \(({\cal D}[\rho_p])_{12}=-\Gamma_{\rm PD }\rho_{12}\) and therefore
models the relaxation/dephasing of the pseudo-Dirac off-diagonal coherence}.

The Majoron-induced term is not a dissipator. It is a Hamiltonian perturbation produced by the time-dependent background $\dot\phi/f_{\rm eff}$. The dissipator instead describes the coarse-grained loss of coherence or relaxation toward the instantaneous response. For the off-diagonal coherence channel, we parameterize
\begin{equation}
{\cal D}[\rho_p]_{12}
=
-\Gamma_{\rm PD}(p)\rho_{12}(p).
\end{equation}
The off-diagonal coherence then obeys
\begin{equation}
\dot\rho_{12}(p,t)
+
\left[
\Gamma_{\rm PD}(p)+i\Delta E_p
\right]\rho_{12}(p,t)
=
B_p(t),
\label{rho12_eq}
\end{equation}
where \(B_p(t)\) is the Majoron-induced Hamiltonian drive projected onto the off-diagonal channel. To leading order around a diagonal
occupation background, \(B_p(t)\) is proportional to
\(g_p(t)D_p\), with \(D_p\equiv\rho_{11}(p)-\rho_{22}(p)\). Hence
\(B_p(t)\) is controlled by \(\dot\phi\), and its time variation is
controlled by \(\ddot\phi\).

Equation~\eqref{rho12_eq} displays the distinct roles of the relevant physical scales. The splitting \(\Delta E_p\) generates coherent phase precession, the rate \(\Gamma_{\rm PD}(p)\) controls relaxation or decoherence, and the drive \(B_p(t)\) is sourced by the evolving Majoron background. Freezing requires the coexistence of coherence and finite relaxation: without coherence there is no phase-lagged pseudo-Dirac current response, while in the absence of relaxation or coarse graining the dynamics remains essentially unitary and the exchange is reversible rather than an effective irreversible response.

\subsection{Lag coherence}

It is useful to define the instantaneous coherence preferred by the drive,
\begin{equation}
\rho_{12}^{\rm inst}(p,t)
=
\frac{B_p(t)}
{\Gamma_{\rm PD}(p)+i\Delta E_p}.
\end{equation}
The actual coherence does not follow this instantaneous value exactly. The lag coherence is defined as
\begin{equation}
\chi_p(t)
=
\rho_{12}^{\rm inst}(p,t)-\rho_{12}(p,t).
\end{equation}
Using Eq.~\eqref{rho12_eq}, one obtains a relaxation equation for $\chi_p$ of the schematic form
\begin{equation}
\dot\chi_p+
\left[
\Gamma_{\rm PD}(p)+i\Delta E_p
\right]\chi_p
=
\dot\rho_{12}^{\rm inst}(p,t).
\end{equation}

Since \(B_p(t)\) is controlled by \(\dot\phi\), the source term
\(\dot\rho^{\rm inst}_{12}\) is controlled by \(\ddot\phi\). This microscopic relation anticipates the effective source term in the local lag equation, Eq.~\eqref{eq:local-lag-ddotphi}, derived below.

Writing
\begin{equation}
\rho_{12}(p,t)=R_p(t)+iI_p(t),
\end{equation}
the imaginary quadrature is the component selected by the  pseudo-Dirac number density operator in the phase convention used here. The macroscopic lag variable is obtained by summing the lagged coherence selected by the pseudo-Dirac number-density operator over the hidden ensemble:
\begin{equation}
X(t)
=
\int\frac{d^3p}{(2\pi)^3}
\,2E_p a_p\,{\rm Im}\,\chi_p(t),
\label{X_micro}
\end{equation}
where $a_p$ denotes the transition matrix element of \(J^0_{\rm PD}\). Equation~\eqref{X_micro} is the microscopic interpretation of the collective lag variable used in the effective response theory.

This construction also clarifies why \(X\) is not an energy density. The
physical hidden density is built from diagonal occupations, whereas \(X\)
is built from off-diagonal lag coherence. The response-weighted density
\(\rho_N^{\rm resp}\) is therefore not an additional gravitating component, but a measure of how efficiently the physical hidden reservoir participates in the coherent lag response.

\section{Retarded response and effective freezing equation}

\subsection{Causal response kernel}

The hidden reservoir has finite memory. Its response to the Majoron drive therefore depends on the recent history of the scalar background, not only on its instantaneous value. We describe the collective lag variable by a causal retarded kernel,
\begin{equation}
X(t)
=
\int_{-\infty}^{t}dt'\,
K(t-t')F(t') ,
\label{eq:causal-response-kernel}
\end{equation}
where \(F(t)\) is the effective drive and \(K(t-t')\) vanishes for \(t<t'\). The upper limit enforces causality.

The exponential kernel used below should be understood as a leading effective-pole reduction of the full microscopic response, not as a fundamental ansatz. At the level of a fixed pseudo-Dirac momentum mode, the lagged coherence variable obeys schematically
\begin{equation}
\dot\chi_p+\Lambda_p\chi_p=S_p(t),
\qquad
\Lambda_p\equiv \Gamma_p+i\Delta E_p ,
\label{eq:mode-lag-equation}
\end{equation}
where \(\Gamma_p\) is the relaxation or dephasing rate and \(\Delta E_p\) is the pseudo-Dirac precession frequency. The causal solution is
\begin{equation}
\chi_p(t)
=
\int_{-\infty}^{t}dt'\,
e^{-\Lambda_p(t-t')}\,
S_p(t') .
\label{eq:mode-lag-solution}
\end{equation}
After projection onto the quadrature selected by the pseudo-Dirac current and summation over the hidden ensemble, the collective kernel is therefore generally a mode-summed, damped-oscillatory kernel. We may write it in the schematic form
\begin{equation}
\begin{aligned}
K(\tau)
&=
\Theta(\tau)
\int\frac{d^3p}{(2\pi)^3}\,
{\cal W}_p\,
e^{-\Gamma_p\tau}
\bigl[
A_p\cos(\Delta E_p\tau)
\\
&\qquad\qquad
+
B_p\sin(\Delta E_p\tau)
\bigr],
\qquad
\tau\equiv t-t' .
\end{aligned}
\label{eq:general-mode-kernel}
\end{equation}
Here \({\cal W}_p\), \(A_p\), and \(B_p\) encode the response weight, current projection, occupation factors, and the phase convention of the microscopic response.

We now use the cold nonrelativistic reservoir approximation summarized
in Eq.~\eqref{eq:cold-nr-splitting}. If the response-weighted support also has weak
variation in the effective relaxation rate and in the pseudo-Dirac
precession frequency, the mode-dependent quantities may be replaced by
their representative values,
\[
\Gamma_p\simeq\Gamma_{\rm PD},
\qquad
\Delta E_p\simeq\Delta E_\ast .
\]
The remaining momentum dependence fixes the effective residue of the
collective response.

The reduction of the full damped-oscillatory kernel to a single
exponential form should be understood as a low-frequency effective-pole
approximation, not as an equality of the microscopic kernel in the time
domain. Equivalently, the effective residue \(\beta\) is defined by
matching the zero-frequency response, or zeroth moment, of the full
retarded kernel. The detailed matching is given in
Appendix~\ref{app:effective-pole-reduction}. The resulting effective
kernel is
\begin{equation}
K_{\rm eff}(\tau)
\simeq
\beta e^{-\Gamma_{\rm PD}\tau}\Theta(\tau),
\label{eq:effective-exponential-kernel}
\end{equation}
where \(\beta\) is the corresponding low-frequency residue. It absorbs
the current projection, occupation factors, phase convention, and the
order-one dependence on \(\Delta E_\ast/\Gamma_{\rm PD}\). Corrections
are controlled by the response-weighted momentum spread, by the spread
of \(\Gamma_p\), by coefficient drift over the memory time, and by the
ratio of the macroscopic source frequency to the microscopic relaxation
and precession scales.

Substituting Eq.~\eqref{eq:effective-exponential-kernel} into Eq.~\eqref{eq:causal-response-kernel} gives
\begin{equation}
X(t)
=
\beta
\int_{-\infty}^{t}dt'\,
e^{-\Gamma_{\rm PD}(t-t')}F(t') .
\label{eq:X-exponential-integral}
\end{equation}
Differentiating with respect to time yields
\begin{equation}
\dot X+\Gamma_{\rm PD}X=\beta F .
\label{eq:local-lag-general-F}
\end{equation}
The rate \(\Gamma_{\rm PD}\) is the effective relaxation rate of the hidden reservoir response. It should not be identified with the macroscopic freezing rate of the Majoron velocity.

The Majoron-induced perturbation of the hidden sector is proportional to \(\dot\phi/f_{\rm eff}\). The lag variable measures the delayed failure of the reservoir to follow the changing instantaneous response. Therefore the source for the lag is the time derivative of the instantaneous perturbation. At leading order in the slowly drifting reservoir coefficients, this gives
\begin{equation}
F(t)=\ddot\phi(t),
\label{eq:F-ddotphi}
\end{equation}
and Eq.~\eqref{eq:local-lag-general-F} becomes
\begin{equation}
\dot X+\Gamma_{\rm PD}X=\beta\ddot\phi .
\label{eq:local-lag-ddotphi}
\end{equation}
Equation~\eqref{eq:local-lag-ddotphi} is not a phenomenological friction law. It is the local short-memory form of a retarded microscopic response. The Majoron-induced drive does not act as a direct local friction term; it first generates a lagged hidden response \(X\), which is subsequently matched onto the energy transfer \(Q\).

\subsection{Energy transfer}

We work in a spatially flat Friedmann--Robertson--Walker spacetime,
\begin{equation}
ds^2=-dt^2+a^2(t)d{\bf x}^2,
\end{equation}
and treat the Majoron as a homogeneous scalar field $\phi(t)$. Its energy density and pressure are
\begin{equation}
\rho_\phi=\frac12\dot\phi^2+V(\phi),
\qquad
p_\phi=\frac12\dot\phi^2-V(\phi).
\end{equation}
In the presence of energy exchange between the Majoron sector and the hidden reservoir, the individual stress tensors are not separately conserved:
\begin{equation}
\nabla_\mu T_\phi^{\mu\nu}=-Q^\nu,
\qquad
\nabla_\mu T_h^{\mu\nu}=Q^\nu .
\end{equation}
The total stress tensor remains conserved,
\begin{equation}
\nabla_\mu\left(T_\phi^{\mu\nu}+T_h^{\mu\nu}\right)=0.
\end{equation}
Projecting along the comoving cosmological four-velocity gives
\begin{equation}
\dot\rho_\phi+3H(\rho_\phi+p_\phi)=-Q ,
\end{equation}
where $Q>0$ denotes energy transfer from the Majoron sector to the hidden reservoir. Using the expressions for $\rho_\phi$ and $p_\phi$, we obtain
\begin{equation}
\dot\phi
\left(
\ddot\phi+3H\dot\phi+V_\phi
\right)
=-Q .
\end{equation}
On a nonstatic branch with $\dot\phi\neq 0$, this may be written as
\begin{equation}
\ddot\phi+3H\dot\phi+V_\phi
=
-\frac{Q}{\dot\phi}.
\label{scalar_nonstatic}
\end{equation}
The fundamental equation is the undivided energy-balance equation. The divided form in Eq.~\eqref{scalar_nonstatic} is used only on a branch with nonzero but possibly very small velocity.

\subsection{Freezing equation}

The collective lag variable \(X\) represents the coarse-grained delayed response of the hidden pseudo-Dirac reservoir. The scalar-sector energy-transfer rate must therefore vanish in the absence of a lagged response,
\begin{equation}
X=0
\qquad\Rightarrow\qquad
Q=0 .
\label{eq:Q-vanishes-without-lag}
\end{equation}
In the local small-lag regime, the energy-transfer closure can be expanded as an analytic function of \(X\),
\begin{equation}
Q(X)
=
\alpha X+\alpha_2 X^2+\alpha_3 X^3+\cdots ,
\label{eq:Q-small-lag-expansion}
\end{equation}
where the coefficients encode the normalization, current projection, occupation factors, and sign convention of the hidden-sector response. These coefficients may depend slowly on the cosmological background and on the reservoir state, but they are treated as constant over the short memory time of the response.

The present work keeps only the leading term in the small-lag linear-response regime,
\begin{equation}
Q\simeq \alpha X .
\label{eq:Q-alpha-X}
\end{equation}
Higher powers of \(X\) describe nonlinear response corrections and are neglected in the leading effective description. This closure is not an independent friction ansatz. The nonlocality of the hidden-sector response has already been encoded in the retarded evolution of \(X\); Eq.~\eqref{eq:Q-alpha-X} is only the leading local matching between the lag variable and the scalar-sector energy-transfer rate.

In the Markovian regime,
\begin{equation}
|\dot X|\ll \Gamma_{\rm PD}|X| ,
\label{eq:Markovian-X-condition}
\end{equation}
the lag equation gives
\begin{equation}
X\simeq
\frac{\beta}{\Gamma_{\rm PD}}\ddot\phi .
\label{eq:X-Markovian-ddotphi}
\end{equation}
Hence
\begin{equation}
Q\simeq q_{\rm exch}\ddot\phi,
\qquad
q_{\rm exch}\equiv
\frac{\alpha\beta}{\Gamma_{\rm PD}} .
\label{eq:Q-qexch-ddotphi}
\end{equation}
The coefficient \(q_{\rm exch}\) is not a relaxation rate. The microscopic relaxation rate is \(\Gamma_{\rm PD}\), which controls the memory time of the hidden reservoir. By contrast, \(q_{\rm exch}\) is an effective exchange coefficient that measures how the Markovian remnant of the lagged reservoir response contributes to the scalar-sector energy transfer.

Substituting Eq.~\eqref{eq:Q-qexch-ddotphi} into the scalar energy-balance equation gives
\begin{equation}
\dot\phi
\left(
\ddot\phi+3H\dot\phi+V_\phi
\right)
+
q_{\rm exch}\ddot\phi
=
0 .
\label{eq:regular-freezing-equation}
\end{equation}
This is the regular form of the effective freezing equation. On a nonstatic branch with \(\dot\phi\neq 0\), Eq.~\eqref{eq:regular-freezing-equation} can be rewritten as
\begin{equation}
\ddot\phi+3H\dot\phi+V_\phi
+
q_{\rm exch}\frac{\ddot\phi}{\dot\phi}
=
0 .
\label{eq:divided-freezing-equation}
\end{equation}
Equation~\eqref{eq:divided-freezing-equation} is the convenient force-like form of the effective scalar equation on the nonstatic branch. The term proportional to \(q_{\rm exch}\ddot\phi/\dot\phi\) is the Markovian remnant of the delayed hidden response. It should not be interpreted as an externally inserted local friction force.

The appearance of \(\dot\phi\) in the denominator of Eq.~\eqref{eq:divided-freezing-equation} does not signal a fundamental singularity. It is only a consequence of dividing the regular energy-balance equation by \(\dot\phi\). Therefore the limit \(\dot\phi\to 0\) must be discussed using Eq.~\eqref{eq:regular-freezing-equation}, not the divided form Eq.~\eqref{eq:divided-freezing-equation}. If \(\dot\phi=0\) and \(\ddot\phi=0\) exactly, Eq.~\eqref{eq:regular-freezing-equation} is automatically satisfied at the level of scalar energy balance, since there is no kinetic energy flow and \(Q\simeq q_{\rm exch}\ddot\phi\) also vanishes. However, this does not imply that an arbitrary point on the potential is a complete static solution of the dynamical system. In the absence of reservoir-supported motion, an exactly static configuration must also satisfy the usual force-balance condition \(V_\phi=0\).

Away from a potential extremum, the relevant solution is therefore not an exactly static one, but a response-dominated quasi-static branch. In this branch the velocity remains nonzero but very small,
\begin{equation}
0<|\dot\phi|\ll q_{\rm exch}.
\label{eq:nonstatic-freezing-branch}
\end{equation}
The divided equation Eq.~\eqref{eq:divided-freezing-equation} is then a valid local description of the nonstatic branch, while the undivided equation Eq.~\eqref{eq:regular-freezing-equation} provides the regular continuation toward the static limit. This is the sense in which the hidden reservoir freezes the Majoron motion: it dynamically suppresses \(|\dot\phi|\) without requiring the Majoron to sit exactly at a potential extremum.

\section{Cosmological freezing and consistency window}

\subsection{Response-dominated freezing branch}

The freezing branch corresponds to the regime in which the exchange term in Eq.~\eqref{eq:divided-freezing-equation} dominates the acceleration balance and dynamically suppresses the scalar velocity. The relevant conditions are
\begin{equation}
|\dot\phi|\ll q_{\rm exch},
\qquad
3H|\dot\phi|\ll |V_\phi|.
\label{freezing_conditions}
\end{equation}
It is useful to compare the second condition in Eq.~\eqref{freezing_conditions} with the ordinary slow-roll estimate. If the hidden response is switched off, \(Q=0\), and the scalar acceleration is neglected in the usual slow-roll approximation, the Majoron equation gives
\begin{equation}
3H\dot\phi_{\rm SR}+V_\phi\simeq 0,
\qquad
\dot\phi_{\rm SR}\simeq -\frac{V_\phi}{3H}.
\label{eq:ordinary-slow-roll-velocity}
\end{equation}
Therefore the second freezing condition can be written as
\begin{equation}
3H|\dot\phi|\ll |V_\phi|
\qquad\Longleftrightarrow\qquad
|\dot\phi|\ll |\dot\phi_{\rm SR}| .
\label{eq:freezing-vs-slow-roll-velocity}
\end{equation}
This shows that the response-dominated freezing branch is not an ordinary slow-roll branch. The Majoron velocity is required to be much smaller than the velocity that would be obtained from Hubble friction alone. In this sense, the second condition in Eq.~\eqref{freezing_conditions} states that ordinary Hubble damping is subdominant in the force balance. The velocity suppression is instead provided by the exchange term generated by the delayed hidden-reservoir response.

Consequently, once the cosmological evolution has already brought the field into a sub-slow-roll regime, $|\dot\phi|\ll |\dot\phi_{\rm SR}|$, the second freezing condition is automatically satisfied. The remaining nontrivial requirement for the onset of response-dominated freezing is then the first condition in Eq.~\eqref{freezing_conditions}, $|\dot\phi|\ll q_{\rm exch}$.
The role of the hidden pseudo-Dirac reservoir is to maintain and further stabilize this sub-slow-roll state even when the intrinsic Majoron curvature scale is larger than the Hubble scale.

These are not ordinary slow-roll conditions, but response-dominated freezing conditions. The first condition ensures that the exchange term is dynamically important compared with the bare acceleration term. The second condition states that ordinary Hubble damping is not the primary source of the velocity suppression.

Let
\begin{equation}
u\equiv \dot\phi .
\end{equation}
In the freezing regime, Eq.~\eqref{eq:divided-freezing-equation} approximately gives
\begin{equation}
\dot u\simeq -\lambda_{\rm fr}u,
\qquad
\lambda_{\rm fr}
\equiv
\frac{|V_\phi|}{q_{\rm exch}},
\label{lambda_fr}
\end{equation}
where the sign convention is chosen so that $\lambda_{\rm fr}>0$ corresponds to damping of the Majoron velocity. The solution is
\begin{equation}
u(t)\simeq u_i
\exp\left[
-\int_{t_i}^{t}dt'\,\lambda_{\rm fr}(t')
\right].
\end{equation}
Thus the hidden reservoir response can dynamically suppress the Majoron velocity. When the kinetic energy is much smaller than the potential energy,
\begin{equation}
\frac12\dot\phi^2\ll V(\phi),
\end{equation}
the equation of state becomes
\begin{equation}
w_\phi
=
\frac{\frac12\dot\phi^2-V(\phi)}
{\frac12\dot\phi^2+V(\phi)}
\simeq -1.
\end{equation}
The resulting state is not an exactly static cosmological constant, but a metastable nonequilibrium frozen phase maintained by the delayed response of the hidden reservoir.

\subsection{Response density and exchange coefficient}

The physical hidden sterile energy density is determined by the
diagonal occupations of the hidden pseudo-Dirac modes,
\begin{equation}
\rho_N =
\int \frac{d^3p}{(2\pi)^3}\,E_p N_p,
\qquad
N_p=\rho_{11}(p)+\rho_{22}(p).
\end{equation}
This is the actual energy density carried by the hidden pseudo-Dirac
reservoir and is the quantity that contributes to the gravitational
stress-energy tensor. However, the response strength relevant for
freezing is not controlled by \(\rho_N\) alone. It also depends on the
effective susceptibility of the hidden ensemble to the lagged
pseudo-Dirac current response, namely on how coherently and efficiently
the physical hidden density is projected onto the response channel.

We therefore introduce a response-weighted density
\begin{equation}
\rho_N^{\rm resp}=C_{\rm resp}\rho_N .
\end{equation}
Here \(C_{\rm resp}\) is a dimensionless response-enhancement factor.
It should not be interpreted as a population fraction of the hidden
reservoir and is not required to satisfy \(0<C_{\rm resp}\leq 1\).
Rather, it plays the role of an effective coherence or susceptibility
quality factor, measuring how strongly the physical hidden-sector
density \(\rho_N\) participates in the lagged pseudo-Dirac current
response. Thus \(\rho_N^{\rm resp}\) is a response-weighted bookkeeping
density entering the Majoron exchange term, not an independently
gravitating matter density.

The parametric origin of \(C_{\rm resp}\) can be traced to the
mode-level response. For a diagonal occupation background, the
Majoron-induced off-diagonal drive scales as
\[
B_p(t)\sim a_p D_p\frac{\dot\phi(t)}{f_{\rm eff}},
\qquad
D_p\equiv \rho_{11}(p)-\rho_{22}(p)\equiv \eta_p N_p .
\]
The instantaneous coherence is therefore
\[
\rho^{\rm inst}_{12}(p,t)
\sim
\frac{B_p(t)}{\Gamma_{\rm PD}+i\Delta E_p}.
\]
Since the lag variable measures the delayed failure of the reservoir to
follow this instantaneous response, the lag coherence is sourced by
\(\dot\rho^{\rm inst}_{12}\). In the short-memory regime this gives,
parametrically,
\[
\chi_p
\sim
\frac{a_pD_p}{f_{\rm eff}}
\frac{\ddot\phi}{(\Gamma_{\rm PD}+i\Delta E_p)^2}.
\]
For the representative response modes with
\(\Delta E_p\simeq \Delta E_*\), order-one phase factors are absorbed
into the effective response coefficients. The collective lag variable then scales as
\[
X \sim
\frac{a_{\rm eff}^2\eta_D\rho_N}
{f_{\rm eff}(\Gamma_{\rm PD}^2+\Delta E_*^2)}
\ddot\phi .
\]
For the representative response regime \(\Delta E_*\sim\Gamma_{\rm PD}\),
this reduces, up to order-one factors, to
\[
X \sim
\frac{a_{\rm eff}^2\eta_D\rho_N}
{f_{\rm eff}\Gamma_{\rm PD}^2}
\ddot\phi .
\]
Matching \(Q=\alpha X\) gives
\[
q_{\rm exch}
\sim
\frac{\alpha a_{\rm eff}^2\eta_D\rho_N}
{f_{\rm eff}\Gamma_{\rm PD}^2}.
\]

The exchange coefficient is controlled by this coherently responding
part of the hidden reservoir. Parametrically,
\begin{equation}
q_{\rm exch}\sim
\frac{\rho_N^{\rm resp}}{m_N f_{\rm eff}}
=
\frac{C_{\rm resp}\rho_N}{m_N f_{\rm eff}} .
\end{equation}
We parameterize the local energy-transfer matching coefficient as
\(\alpha=\xi_{\rm match}m_N\). This does not introduce an additional
relaxation rate; the relaxation dynamics is already contained in the
retarded susceptibility through \(\Gamma_{\rm PD}\). Rather, \(\alpha\)
converts the energy-density-normalized lag variable \(X\) into the
macroscopic transfer term \(Q=\alpha X\). For a cold nonrelativistic
reservoir, the natural microscopic energy scale in this matching is
\(E_p\simeq m_N\).
The response-enhancement factor is estimated as
\begin{equation}
C_{\rm resp}
=
\xi_{\rm match} a_{\rm eff}^2\eta_D
\left(
\frac{m_N}{\Gamma_{\rm PD}}
\right)^2.
\qquad
\xi_{\rm match} \equiv\frac{\alpha}{m_N} .
\end{equation}
Thus \(C_{\rm resp}\) is not an independent matter fraction. It is a
dimensionless response factor that collects the local energy-transfer
normalization $\xi_{\rm match}$, the two current-projection factors
\(a_{\rm eff}^2\), the available pseudo-Dirac imbalance \(\eta_D\), and
the finite-memory enhancement measured by $(m_N/\Gamma_{\rm PD})^2$.
Order-one factors depending on \(\Delta E_\ast/\Gamma_{\rm PD}\) and on
the detailed response-weighted momentum distribution are absorbed into
\(a_{\rm eff}\) and \(\xi_{\rm match}\).

\subsection{Markovian consistency}

The local lag equation is valid only in the short-memory regime. The Markovian approximation requires
\begin{equation}
\Gamma_{\rm PD}\gg
\left|
\frac{d}{dt}\ln F
\right|,
\qquad
F=\ddot\phi .
\end{equation}
A conservative cosmological consistency condition is
\begin{equation}
\Gamma_{\rm PD}\gg
\max(H,m_{\phi},\lambda_{\rm fr}).
\label{markov_condition}
\end{equation}
This hierarchy means that the hidden reservoir memory time is short compared with the macroscopic timescales over which the Majoron background changes. At the same time, the exchange coefficient must remain large enough to satisfy the freezing conditions in Eq.~\eqref{freezing_conditions}. The mechanism therefore requires a controlled balance: the reservoir must respond slowly enough to generate a lag, but fast enough for the Markovian reduction to be valid.

As an illustrative hierarchy, one may consider
\begin{equation}
\Gamma_{\rm PD}\gg m_{\phi}\gg H_0.
\end{equation}
In this case the intrinsic Majoron mass may be larger than the Hubble scale, while the hidden reservoir response remains sufficiently fast to justify a local effective description. If the relaxation rate evolves and eventually becomes comparable to the macroscopic variation rate, the Markovian description will break down and the full nonlocal kernel must be retained. Such a transition would correspond to a future non-Markovian regime, not necessarily to an immediate failure of the mechanism.

\subsection{Benchmark consistency window}

We now illustrate that the response-induced frozen branch can be realized
within a broad parametric window. For the benchmark charge assignment
\(q_1=2\), \(q_2=13\), with
\(v_1\simeq 2.2\times 10^{11}\,{\rm GeV}\) and
\(v_2\simeq 1.0\times 10^9\,{\rm GeV}\), one obtains
\(f_{\rm eff}\simeq 5.0\times 10^8\,{\rm GeV}\). The leading
Planck-suppressed operator then gives a dark-energy-scale potential
\(\Lambda_\phi^4\simeq 2.6\times 10^{-11}|\kappa|\,{\rm eV}^4\)
and an intrinsic Majoron mass
\(m_\phi\simeq 10^{-23}\sqrt{|\kappa|}\,{\rm eV}\).

As a representative hidden-reservoir choice, take
\(m_N=10^{-3}\,{\rm eV}\), \(\mu_h=10^{-8}\,{\rm eV}\), and
\(\Gamma_{\rm PD}\simeq 2\mu_h=2\times 10^{-8}\,{\rm eV}\).
Then \(\Delta E_*\simeq 2\mu_h\simeq \Gamma_{\rm PD}\), and the
representative response regime \(\Delta E_*\sim \Gamma_{\rm PD}\) is
naturally realized. Moreover,
\[
\frac{\Gamma_{\rm PD}}{m_\phi}\sim 2\times 10^{15},
\]
so the short-memory hierarchy \(\Gamma_{\rm PD}\gg m_\phi\gg H_0\) is
comfortably satisfied.

The freezing condition is controlled not by the total hidden density alone,
but by the response-weighted combination entering
\[
q_{\rm exch}\sim \frac{C_{\rm resp}\rho_N}{m_N f_{\rm eff}} .
\]
Thus the required benchmark condition may be expressed as a lower bound on
the response-weighted hidden density,
\[
C_{\rm resp}\rho_N \gg m_N f_{\rm eff} u_{\rm SR},
\qquad
u_{\rm SR}\sim \frac{|V_\phi|}{3H}.
\]
For the benchmark values above, the response factor can be enhanced by the
finite-memory susceptibility, while the actual hidden density \(\rho_N\)
remains a model-dependent input. A complete abundance and perturbation
analysis of the hidden reservoir is left for future work.

\subsection{Finite-memory non-Markovian freezing}

The local Markovian description is not the most general form of the freezing mechanism. When the reservoir memory time is not negligible compared with the macroscopic time scale of the Majoron background, the lag variable \(X\) should not be eliminated through the local relation \(X\simeq (\beta/\Gamma_{\rm PD})\ddot\phi\). Instead, the finite-memory variable \(X(t)\) must be kept as an independent reservoir response determined by the causal kernel.

Using $u=\dot\phi$, 
the regular scalar energy-balance equation can then be written as
\begin{equation}
u\left(\dot u+3Hu+V_\phi\right)+\alpha X=0 .
\label{eq:secV-nonmark-energy-balance}
\end{equation}
This equation is the appropriate starting point for the finite-memory regime. In this form no local Markovian exchange coefficient \(q_{\rm exch}\) has been introduced.

A non-Markovian frozen branch is established when the finite-memory lag response supplies the leading counterterm to the potential-force work,
\begin{equation}
uV_\phi+\alpha X\simeq 0 .
\label{eq:secV-nonmark-leading-balance}
\end{equation}
The inertial and Hubble-drag energy terms must remain subleading,
\begin{equation}
|u\dot u|\ll |uV_\phi|,
\qquad
|3Hu^2|\ll |uV_\phi| .
\label{eq:secV-nonmark-subleading}
\end{equation}
On a nonstatic branch away from a potential extremum, \(uV_\phi\neq 0\), these conditions may be written as
\begin{equation}
{\cal R}_{\rm acc}
\equiv
\left|\frac{\dot u}{V_\phi}\right|
\ll 1,
\qquad
{\cal R}_{H}
\equiv
\left|\frac{3Hu}{V_\phi}\right|
\ll 1 .
\label{eq:secV-nonmark-subleading-ratios}
\end{equation}
Thus the finite-memory frozen branch is characterized by the leading response balance in Eq.~\eqref{eq:secV-nonmark-leading-balance}, together with the smallness of the inertial and Hubble-drag ratios in Eq.~\eqref{eq:secV-nonmark-subleading-ratios}. This criterion does not require eliminating \(X\) in favor of a local Markovian exchange coefficient.

The usual Markovian freezing conditions are recovered as a limiting case. If the hidden response becomes local,
\begin{equation}
X\simeq \frac{\beta}{\Gamma_{\rm PD}}\dot u ,
\label{eq:secV-nonmark-markovian-limit}
\end{equation}
then the leading balance in Eq.~\eqref{eq:secV-nonmark-leading-balance} becomes
\begin{equation}
q_{\rm exch}\dot u+uV_\phi\simeq 0,
\qquad
q_{\rm exch}\equiv \frac{\alpha\beta}{\Gamma_{\rm PD}} .
\label{eq:secV-nonmark-markovian-balance}
\end{equation}
Therefore
\begin{equation}
\dot u\simeq -\frac{uV_\phi}{q_{\rm exch}} .
\label{eq:secV-nonmark-u-dot-markovian}
\end{equation}
In this limit, the Markovian hierarchy \(|u|\ll q_{\rm exch}\) implies
\begin{equation}
|u\dot u|
\simeq
\left|\frac{u^2V_\phi}{q_{\rm exch}}\right|
\ll
|uV_\phi| ,
\label{eq:secV-nonmark-inertial-from-markovian}
\end{equation}
while the Markovian condition \(3H|u|\ll |V_\phi|\) is identical to the second condition in Eq.~\eqref{eq:secV-nonmark-subleading-ratios}. Thus the two Markovian freezing hierarchies imply the two subleading non-Markovian hierarchies whenever the local reduction is valid. The converse is not required: the finite-memory formulation can define a frozen branch even when \(X\) cannot be eliminated in favor of a local \(q_{\rm exch}\).

In addition to the leading balance and the subleading hierarchy, the
finite-memory branch must satisfy a dynamical stability condition.
The leading balance \(uV_\phi+\alpha X\simeq 0\) must be preserved by
the finite-memory evolution of \(X\). Equivalently, small departures
from the response-dominated branch must damp rather than grow. As shown in
Appendix~E, combining the leading balance with the finite-memory
equation for \(X\) gives an effective velocity equation of the form
\begin{equation}
\dot u\simeq -\lambda_{\rm eff}u .
\end{equation}
The non-Markovian frozen branch is dynamically acceptable when
\begin{equation}
\lambda_{\rm eff}>0 .
\end{equation}
This condition is not an additional freezing hierarchy. Rather, it is
the stability requirement that selects the branch on which the
finite-memory response suppresses the Majoron velocity instead of
amplifying it.

\subsection{Physical interpretation}

The mechanism may be summarized as follows. The Majoron motion drives the hidden pseudo-Dirac reservoir through a derivative coupling. The reservoir has an intrinsic phase clock set by $\Delta E_p$ and a finite relaxation rate $\Gamma_{\rm PD}$. The combination of coherent phase evolution and finite relaxation produces a phase-lagged response. The lagged response is encoded in $X$, and the energy-transfer matching $Q=\alpha X$ allows the hidden reservoir to absorb energy from the Majoron sector on the freezing branch.

If the reservoir were perfectly instantaneous, no lag would be generated. If the reservoir were perfectly coherent but without relaxation, energy would tend to be exchanged reversibly between the Majoron and hidden modes. The useful regime is therefore a nonequilibrium open-system regime with long-lived coherence and finite relaxation. In this regime, the response is neither a simple thermal friction term nor a purely reversible oscillation. It is a retarded collective response capable of suppressing the Majoron velocity.

\section{Discussion and conclusions}

We have proposed a retarded-response freezing mechanism for Majoron dark energy. The central idea is that a physical Majoron need not remain dark-energy-like solely because its potential is ultra-flat or its mass is of order $H_0$. Instead, its motion can be dynamically suppressed by the finite-memory response of a hidden pseudo-Dirac sterile reservoir.

The effective theory contains three key ingredients. First, the Majoron couples derivatively to the hidden pseudo-Dirac current, so that a homogeneous Majoron background drives the hidden reservoir. Second, the hidden reservoir possesses a pseudo-Dirac coherence channel with intrinsic phase precession set by $\Delta E_p$ and relaxation described by $\Gamma_{\rm PD}$. Third, the coarse-grained reservoir response generates a collective lag variable $X$, which determines the energy transfer between the Majoron sector and the hidden reservoir.

In the short-memory regime, the retarded response reduces to the local lag equation
\(\dot X+\Gamma_{\rm PD}X=\beta\ddot\phi\). Together with the matching
relation \(Q=\alpha X\), this gives \(Q\simeq q_{\rm exch}\ddot\phi\).
The resulting scalar equation contains the exchange structure
\(q_{\rm exch}\ddot\phi/\dot\phi\), which is the Markovian remnant of the
delayed hidden response.

The microscopic origin of the lag variable is the phase-lagged off-diagonal coherence of the hidden pseudo-Dirac ensemble. This provides a concrete interpretation of the effective response variable and distinguishes the mechanism from a phenomenological friction term inserted directly into the scalar equation. The response strength is controlled by a response-weighted density $\rho_N^{\rm resp}=C_{\rm resp}\rho_N$, while only the physical density $\rho_N$ contributes directly to the stress-energy tensor.

We do not claim that the pseudo-Dirac sterile reservoir is the unique possible
realization of such a response medium. Rather, it provides a minimal
particle-physics implementation of the required ingredients: a hidden
nonequilibrium reservoir, a sterile two-state system, and a pseudo-Dirac
coherence channel with a small intrinsic splitting. Other hidden-sector
realizations may produce analogous retarded response effects.

Several issues remain for future work. A more complete microscopic model of the collective response factor $C_{\rm resp}$ would be desirable. The Lindblad-type relaxation used here should also be derived from a more explicit hidden-sector environment or many-body dynamics. Finally, a full cosmological analysis, including perturbations and likelihood constraints, is required before the mechanism can be confronted directly with data. The purpose of the present work is more limited: to establish a consistent nonequilibrium effective framework in which Majoron dark energy can be frozen by the retarded collective response of a hidden pseudo-Dirac reservoir.

\appendix

\section{Scalar potential and projection onto the physical Majoron direction}
\label{app:scalar_potential_majoron}

In this appendix we present the scalar potential and show explicitly how the
physical Majoron direction and its periodic potential arise for the charge
assignment specified in Eq.~\eqref{eq:q1q2}. 
The most general renormalizable scalar potential involving the Standard Model
Higgs doublet \(H\) and the two singlet scalars \(\Phi_1,\Phi_2\), consistent
with the gauged \(U(1)_{B-L}\) symmetry, is
\begin{align}
V_{\rm ren}
=&
-\mu_H^2 H^\dagger H
-\mu_1^2 \Phi_1^\dagger\Phi_1
-\mu_2^2 \Phi_2^\dagger\Phi_2
\nonumber\\
&+
\lambda_H(H^\dagger H)^2
+\lambda_1(\Phi_1^\dagger\Phi_1)^2
+\lambda_2(\Phi_2^\dagger\Phi_2)^2
\nonumber\\
&+
\lambda_{12}(\Phi_1^\dagger\Phi_1)(\Phi_2^\dagger\Phi_2)
+\lambda_{H1}(H^\dagger H)(\Phi_1^\dagger\Phi_1)
\nonumber\\
&+\lambda_{H2}(H^\dagger H)(\Phi_2^\dagger\Phi_2).
\label{eq:app_Vren}
\end{align}
For the benchmark charges in Eq.~\eqref{eq:q1q2}, no renormalizable
phase-sensitive operator involving both \(\Phi_1\) and \(\Phi_2\) is allowed
by \(U(1)_{B-L}\).  Thus \(V_{\rm ren}\) fixes the radial vacuum expectation
values but does not generate a potential for the physical Majoron direction.

After electroweak and \(U(1)_{B-L}\) symmetry breaking,
\begin{equation}
H=
\frac{1}{\sqrt2}
\begin{pmatrix}
0\\ v+h
\end{pmatrix},
\qquad
\Phi_i=
\frac{1}{\sqrt2}(v_i+\rho_i)e^{i\theta_i},
\label{eq:app_ssb_fields}
\end{equation}
the stationary conditions are
\begin{align}
\mu_H^2
&=
\lambda_H v^2
+\frac12\lambda_{H1}v_1^2
+\frac12\lambda_{H2}v_2^2,
\nonumber\\
\mu_1^2
&=
\lambda_1 v_1^2
+\frac12\lambda_{12}v_2^2
+\frac12\lambda_{H1}v^2,
\nonumber\\
\mu_2^2
&=
\lambda_2 v_2^2
+\frac12\lambda_{12}v_1^2
+\frac12\lambda_{H2}v^2.
\label{eq:app_tadpole_conditions}
\end{align}
In the CP-even scalar basis \((h,\rho_1,\rho_2)\), the tree-level mass matrix
is
\begin{equation}
\mathcal M^2_{\rm even}
=
\begin{pmatrix}
2\lambda_H v^2 & \lambda_{H1}vv_1 & \lambda_{H2}vv_2\\
\lambda_{H1}vv_1 & 2\lambda_1v_1^2 & \lambda_{12}v_1v_2\\
\lambda_{H2}vv_2 & \lambda_{12}v_1v_2 & 2\lambda_2v_2^2
\end{pmatrix}.
\label{eq:app_even_mass_matrix}
\end{equation}
The CP-even radial modes are assumed to be heavy compared with the late-time
Majoron dynamics and are integrated out in the low-energy effective theory.

A small potential for the physical Majoron can be generated by a
Planck-suppressed operator that is invariant under the gauged
\(U(1)_{B-L}\) symmetry but explicitly breaks the accidental global phase
symmetry of the renormalizable singlet sector.  A gauge-invariant operator
aligned with the physical Majoron direction is
\begin{equation}
V_{\rm br}
=
-
\left[
\kappa
\frac{\Phi_1^{m}(\Phi_2^\dagger)^{n}}
{M_{\rm Pl}^{m+n-4}}
+\mathrm{h.c.}
\right].
\label{eq:app_Vbr_operator}
\end{equation}
This operator is gauge invariant because
\begin{equation}
m q_1- n q_2=0.
\label{eq:app_operator_gauge_invariance}
\end{equation}
Substituting Eq.~\eqref{eq:app_ssb_fields} into
Eq.~\eqref{eq:app_Vbr_operator} and integrating out the radial modes gives
\begin{align}
V_{\rm br}(\phi)
&=
-
\kappa
\frac{1}{M_{\rm Pl}^{m+n-4}}
\left(\frac{v_1}{\sqrt2}\right)^{m}
\left(\frac{v_2}{\sqrt2}\right)^{n}
e^{i(m\theta_1-n\theta_2)}
+\mathrm{h.c.}
\nonumber\\
&=
-
\Lambda_{\phi}^4
\cos\left(\frac{\phi}{f_{\rm eff}}+\delta\right),
\label{eq:app_Vbr_cosine}
\end{align}
where
\begin{equation}
\delta=\arg\kappa,
\qquad
\Lambda_{\phi}^4
\equiv
\frac{2|\kappa|}{2^{(m+n)/2}}
\frac{v_1^{m}v_2^{n}}
{M_{\rm Pl}^{m+n-4}}.
\label{eq:app_LambdaJ_general}
\end{equation}
Thus the renormalizable scalar potential generates the
\(U(1)_{B-L}\)-breaking vacuum and leaves an uneaten physical phase direction,
while the Planck-suppressed operator in
Eq.~\eqref{eq:app_Vbr_operator} gives the physical Majoron a small periodic
potential.

\section{Origin of the pseudo-Dirac mass structure}
\label{app:hidden_pseudo_dirac_origin}

In the main text we use the minimal realization in which the hidden
pseudo-Dirac fermions are singlets under the gauged \(U(1)_{B-L}\),
\begin{equation}
Q_{B-L}(N_h)=Q_{B-L}(S_h)=0 .
\end{equation}
This choice separates the hidden response reservoir from the ordinary
right-handed neutrinos responsible for the observed active-neutrino
masses. It also avoids direct \(B-L\) gauge interactions of the hidden
reservoir. The Majoron coupling to the hidden sector is therefore not a
minimal gauge interaction induced by a \(B-L\) charge assignment. It is
instead treated as an effective hidden-sector portal to the
pseudo-Dirac number current.

Because \(N_h\) and \(S_h\) are \(B-L\) singlets, the pseudo-Dirac mass
terms used in the main text are gauge invariant already at the
effective-field-theory level. The most general symmetric two-state mass
structure relevant for the hidden reservoir is
\begin{equation}
{\cal L}_{\rm mass}^{h}
=
-m_N N_hS_h
-\frac12 \mu_N N_hN_h
-\frac12 \mu_S S_hS_h
+{\rm h.c.}
\end{equation}
Here \(m_N\) is the dominant Dirac mass, while \(\mu_N\) and \(\mu_S\)
are small Majorana entries. The pseudo-Dirac limit corresponds to
\begin{equation}
|\mu_N|,|\mu_S|\ll m_N .
\end{equation}
In the limit \(\mu_N,\mu_S\to0\), the hidden sector recovers a conserved
hidden pseudo-Dirac number symmetry \(U(1)_{N_h}\), under which \(N_h\)
and \(S_h\) carry opposite hidden charges. The small Majorana entries
therefore softly break this hidden number symmetry and are technically
natural in the usual 't Hooft sense: setting them to zero enhances the
symmetry of the hidden sector.

For the benchmark model used in the main text, we take the
exchange-symmetric limit
\begin{equation}
\mu_N\simeq \mu_S\equiv \mu_h .
\end{equation}
This can be viewed as the minimal limit selected by an approximate
hidden exchange symmetry \(N_h\leftrightarrow S_h\). In this limit the
mass Lagrangian becomes
\begin{equation}
{\cal L}_{\rm mass}^{h}
=
-m_N N_hS_h
-\frac12\mu_h N_hN_h
-\frac12\mu_h S_hS_h
+{\rm h.c.}
\end{equation}
The corresponding mass matrix in the \((N_h,S_h)\) basis is
\begin{equation}
{\cal M}_h=
\begin{pmatrix}
\mu_h & m_N \\
m_N & \mu_h
\end{pmatrix}.
\end{equation}
It is diagonalized by the symmetric and antisymmetric combinations
\begin{equation}
N_{1h}=\frac{N_h+S_h}{\sqrt2},
\qquad
N_{2h}=\frac{i(N_h-S_h)}{\sqrt2}.
\end{equation}
The factor of \(i\) is a phase convention chosen so that both physical Majorana masses are positive. With this convention,
\(m_{1h}=m_N+\mu_h\) and \(m_{2h}=m_N-\mu_h\). This is the mass structure used in the main text. The corresponding
energy splitting, pseudo-Dirac number current, and two-state current
projection are discussed in Secs.~II~B and III~A.

\section{Effective-pole reduction of the retarded kernel}
\label{app:effective-pole-reduction}

We explain how the damped-oscillatory microscopic response kernel is
reduced to the single-pole kernel used in the main text. Starting from
the mode-summed retarded kernel,
\begin{align}
&K(\tau)= \nonumber\\
&\Theta(\tau)
\int\frac{d^3p}{(2\pi)^3}\,
W_p e^{-\Gamma_p\tau}
\left[
A_p\cos(\Delta E_p\tau)
+
B_p\sin(\Delta E_p\tau)
\right],
\label{eq:app-full-mode-kernel}
\end{align}
we impose the cold nonrelativistic and narrow-support approximation.
Over the response-weighted support,
\begin{equation}
\Gamma_p\simeq \Gamma_{\rm PD},
\qquad
\Delta E_p\simeq \Delta E_\ast .
\label{eq:app-representative-rates}
\end{equation}
The remaining momentum dependence fixes the effective residues
\begin{equation}
A_\ast
\equiv
\int\frac{d^3p}{(2\pi)^3}\,W_pA_p,
\qquad
B_\ast
\equiv
\int\frac{d^3p}{(2\pi)^3}\,W_pB_p .
\label{eq:app-effective-residues}
\end{equation}
Thus
\begin{equation}
K(\tau)
\simeq
\Theta(\tau)e^{-\Gamma_{\rm PD}\tau}
\left[
A_\ast\cos(\Delta E_\ast\tau)
+
B_\ast\sin(\Delta E_\ast\tau)
\right].
\label{eq:app-reduced-oscillatory-kernel}
\end{equation}

The replacement by a single exponential kernel is not an equality of
the full microscopic kernel in the time domain. It is a low-frequency
effective-pole reduction. The zero-frequency response of the full
kernel is
\begin{equation}
\widetilde K(0)
=
\int_0^\infty d\tau\,K(\tau)
=
\frac{
A_\ast\Gamma_{\rm PD}
+
B_\ast\Delta E_\ast
}{
\Gamma_{\rm PD}^2+\Delta E_\ast^2
}.
\label{eq:app-zero-frequency-response}
\end{equation}
We define the effective single-pole kernel by
\begin{equation}
K_{\rm eff}(\tau)
=
\beta e^{-\Gamma_{\rm PD}\tau}\Theta(\tau),
\label{eq:app-effective-single-pole-kernel}
\end{equation}
and match its zeroth moment to that of the full retarded kernel:
\begin{equation}
\int_0^\infty d\tau\,K_{\rm eff}(\tau)
=
\int_0^\infty d\tau\,K(\tau).
\label{eq:app-zeroth-moment-matching}
\end{equation}
Since
\begin{equation}
\int_0^\infty d\tau\,
\beta e^{-\Gamma_{\rm PD}\tau}
=
\frac{\beta}{\Gamma_{\rm PD}},
\label{eq:app-effective-kernel-moment}
\end{equation}
the matching condition gives
\begin{equation}
\frac{\beta}{\Gamma_{\rm PD}}
=
\widetilde K(0).
\label{eq:app-beta-matching-condition}
\end{equation}
Therefore
\begin{equation}
\beta
=
\Gamma_{\rm PD}
\frac{
A_\ast\Gamma_{\rm PD}
+
B_\ast\Delta E_\ast
}{
\Gamma_{\rm PD}^2+\Delta E_\ast^2
}.
\label{eq:app-beta-effective-residue}
\end{equation}
Thus \(\beta\) is an effective low-frequency residue, not a fundamental
microscopic constant. It absorbs the current projection, occupation
factors, phase convention, and the order-one dependence on
\(\Delta E_\ast/\Gamma_{\rm PD}\).

The residues \(A_\ast\) and \(B_\ast\) should be understood as the
effective residues of the kernel acting on the macroscopic source
\(F=\ddot\phi\), not as the bare off-diagonal drive coefficients
proportional to \(\dot\phi\). The instantaneous coherence has already
introduced one inverse microscopic pole,
\((\Gamma_{\rm PD}+i\Delta E_\ast)^{-1}\), into these residues.
Parametrically,
\[
A_\ast,\;B_\ast
=
{\cal O}\!\left(
\frac{a_{\rm eff}^2\eta_D\rho_N}
{f_{\rm eff}\sqrt{\Gamma_{\rm PD}^2+\Delta E_\ast^2}}
\right).
\]
For the representative response regime
\(\Delta E_\ast\sim\Gamma_{\rm PD}\), this gives
\[
\beta
=
{\cal O}\!\left(
\frac{a_{\rm eff}^2\eta_D\rho_N}
{f_{\rm eff}\Gamma_{\rm PD}}
\right),
\qquad
\frac{\beta}{\Gamma_{\rm PD}}
=
{\cal O}\!\left(
\frac{a_{\rm eff}^2\eta_D\rho_N}
{f_{\rm eff}\Gamma_{\rm PD}^2}
\right).
\]
Therefore \(Q=\alpha X\) reproduces the scaling
\[
q_{\rm exch}
\sim
\frac{\alpha a_{\rm eff}^2\eta_D\rho_N}
{f_{\rm eff}\Gamma_{\rm PD}^2}
\]
used in the main text, up to order-one factors depending on
\(\Delta E_\ast/\Gamma_{\rm PD}\) and on the response quadrature.

The reduction is valid when the macroscopic source frequency is small
compared with the microscopic relaxation and precession scales,
\begin{equation}
\omega_\phi
\ll
\Gamma_{\rm PD},
\qquad
\omega_\phi
\ll
\sqrt{\Gamma_{\rm PD}^2+\Delta E_\ast^2}.
\label{eq:app-low-frequency-validity}
\end{equation}
Corrections are controlled by the response-weighted momentum spread,
the spread of \(\Gamma_p\), the coefficient drift over the memory time,
and the ratios
\(\omega_\phi/\Gamma_{\rm PD}\) and
\(\omega_\phi/\sqrt{\Gamma_{\rm PD}^2+\Delta E_\ast^2}\).

\section{Retarded kernel and Markovian expansion}

Starting from the causal response
\begin{equation}
X(t)
=
\beta
\int_{-\infty}^{t}dt'\,
e^{-\Gamma(t-t')}F(t'),
\end{equation}
differentiation gives
\begin{align}
\dot X(t)
&=
\beta F(t)
-\Gamma\beta
\int_{-\infty}^{t}dt'\,
e^{-\Gamma(t-t')}F(t') \nonumber\\
&=
\beta F(t)-\Gamma X(t).
\end{align}
Therefore
\begin{equation}
\dot X+\Gamma X=\beta F.
\end{equation}
For slowly varying $F$, the integral admits the short-memory expansion
\begin{equation}
X(t)
=
\frac{\beta}{\Gamma}
\left[
F(t)
-\frac{\dot F(t)}{\Gamma}
+\frac{\ddot F(t)}{\Gamma^2}
+\cdots
\right].
\end{equation}
The leading Markovian approximation is valid when
\begin{equation}
\Gamma\gg \left|\frac{d}{dt}\ln F\right|.
\end{equation}
In the application considered in the main text, $F=\ddot\phi$ and $\Gamma=\Gamma_{\rm PD}$.

\section{Non-Markovian stability of the finite-memory frozen branch}
\label{app:nonmarkovian-stability}

In this appendix we show how the effective damping rate of the
finite-memory frozen branch follows from the coupled \(u\)-\(X\)
system. We define
\begin{equation}
u\equiv \dot\phi .
\end{equation}
The finite-memory lag variable obeys
\begin{equation}
\dot X+\Gamma_{\rm PD}X=\beta\dot u .
\label{eq:app-nonmark-X-eq}
\end{equation}
In the non-Markovian frozen branch, \(X\) is not eliminated in favor
of a local Markovian coefficient. Instead, the leading balance is
\begin{equation}
uV_\phi+\alpha X\simeq 0 .
\label{eq:app-nonmark-leading-balance}
\end{equation}
This relation states that the finite-memory reservoir response
cancels the potential-force work at leading order.

Taking a time derivative of Eq.~\eqref{eq:app-nonmark-leading-balance}
gives
\begin{equation}
\alpha\dot X
\simeq
-\dot u\,V_\phi-u^2V_{\phi\phi}.
\label{eq:app-nonmark-Xdot}
\end{equation}
Substituting Eqs.~\eqref{eq:app-nonmark-leading-balance} and
\eqref{eq:app-nonmark-Xdot} into Eq.~\eqref{eq:app-nonmark-X-eq},
we obtain
\begin{equation}
-\frac{1}{\alpha}
\left(
V_\phi\dot u+u^2V_{\phi\phi}
\right)
-\frac{\Gamma_{\rm PD}}{\alpha}uV_\phi
=
\beta\dot u .
\end{equation}
Equivalently,
\begin{equation}
\left(\alpha\beta+V_\phi\right)\dot u
+
\Gamma_{\rm PD}uV_\phi
+
u^2V_{\phi\phi}
\simeq 0 .
\label{eq:app-nonmark-u-eq}
\end{equation}
Therefore, on the finite-memory branch,
\begin{equation}
\dot u\simeq -\lambda_{\rm eff}u ,
\end{equation}
with
\begin{equation}
\lambda_{\rm eff}
\simeq
\frac{\Gamma_{\rm PD}V_\phi+uV_{\phi\phi}}
{\alpha\beta+V_\phi}.
\label{eq:app-nonmark-lambda-eff}
\end{equation}
This expression is branch-dependent because \(u\), \(V_\phi\), and
\(\alpha\beta\) carry the sign information of the response channel.
The finite-memory branch corresponds to the sign choice for which
the reservoir response opposes the potential-force work. With this
damping sign convention, the branch is stable when
\begin{equation}
\lambda_{\rm eff}>0 .
\end{equation}

In the response-dominated regime,
\begin{equation}
|\alpha\beta|\gg |V_\phi|,
\end{equation}
and for sufficiently small velocity,
\begin{equation}
|uV_{\phi\phi}|\ll \Gamma_{\rm PD}|V_\phi|,
\end{equation}
Eq.~\eqref{eq:app-nonmark-lambda-eff} reduces to
\begin{equation}
\lambda_{\rm eff}
\simeq
\frac{\Gamma_{\rm PD}V_\phi}{\alpha\beta}.
\end{equation}
Using the Markovian exchange coefficient
\begin{equation}
q_{\rm exch}\equiv \frac{\alpha\beta}{\Gamma_{\rm PD}},
\end{equation}
this becomes
\begin{equation}
\lambda_{\rm eff}
\simeq
\frac{V_\phi}{q_{\rm exch}} .
\end{equation}
With the damping branch chosen so that \(\lambda_{\rm eff}>0\), this
reproduces the Markovian freezing estimate
\begin{equation}
\lambda_{\rm fr}\simeq \frac{|V_\phi|}{q_{\rm exch}} .
\end{equation}
Thus the local Markovian freezing branch is recovered as a limiting
case of the more general finite-memory system.

\end{document}